\documentclass[final,authoryear,5p,times,twocolumn]{elsarticle}
\usepackage{amssymb,amsmath,mathrsfs} 
\usepackage{verbatim}
\usepackage{subfig}

\journal{Journal of Mathematical Psychology}

\begin{document}

\begin{frontmatter}

\title{Properties of the Nonparametric Maximum Likelihood {ROC} Model with a Monotonic Likelihood Ratio}

\author{Lucas Tcheuko}
\address{University of Maryland College Park}
\ead{lucast@umd.edu}
\author{Frank Samuelson}
\address{U.S. Food and Drug Administration\\
10903 New Hampshire Avenue\\
Building 62, Room 3102\\
Silver Spring, MD 20993-0002}

\maketitle
\newcommand{\ud}{\text{d}}

\begin{abstract}
We expect that some observers
in perceptual signal detection experiments, such as radiologists, 
will make rational decisions, and therefore ratings from those
observers are expected to form a convex ROC curve.  However,
measured and published curves are often not convex.  This article
examines the convexity-constrained nonparametric maximum
likelihood estimator of the ROC curve given by Lloyd (2002).
Like Lloyd we use the Pool Adjacent Violator Algorithm
(PAVA) to construct the estimate of the convex curve.  We
present a direct proof that this estimate is a convex hull
of the empirical ROC curve.  The estimate is simple to construct 
by hand, and follows the suggestions by Pesce, et~al.~(2010).

We examine the properties of this constrained nonparametric
maximum likelihood estimator (NPMLE) under a large
number of experimental conditions.  In particular we examine
the behavior of the area under the curve which is often used as summary
metric of diagnostic performance.  This constrained ROC
estimator gives an area under the curve (AUC) estimate that
is biased high with respect to the usual empirical AUC
estimate, but may be less biased with respect to the
underlying continuous true AUC value.  The constrained ROC
estimator has lower variance than the usual empirical one.
Unlike previous authors who used complex bootstrapping to 
estimate the variance of the constrained NPMLE
we demonstrate that standard unbiased estimators of variance 
work well to estimate the variance of the NPMLE AUC.

\begin{keyword} constrained  estimates \sep empirical likelihood \sep PAVA algorithm \sep convex ROC curves
\end{keyword}

\end{abstract}
\end{frontmatter}

\section{Introduction}

In signal detection 
experiments, such as diagnostic medical imaging studies,
an image is presented to an observer who
gives an ordered categorical rating or score that indicates his
confidence in the presence of a signal or disease in that
image.  For example, a subject may give a rating 
of one through five, with five indicating the observer
is very confident that the image contains a signal.
The probability distribution of ratings given to the subjects who 
have disease we call $F$, and the 
distribution of ratings given to the subjects who 
do not have disease we call $H$.  Example densities are shown
in Figure~\ref{histogram}.  Typically we assume that these 
ratings arise from some continuous underlying perceptual distribution 
and are discretized by the observer for convenience.

From this task we can create measures of diagnostic performance.
At each rating category we can calculate the true postive rate (TPR), also called sensitivity, and the 
false positive rate (FPR) or 1-specificity.
The plot of  TPR versus FPR across all ordered categories 
yields an empirical estimate of the 
Receiver Operating Characteristic (ROC) curve~\citep{GreenNSwets}.  An example
of such a curve is shown in Figure~\ref{Chan}. 
The area under an ROC curve (AUC) is the probability that 
the radiologist will give a higher rating to a diseased patient
than a patient without disease and is used to rate the performance 
of the diagnoses.

An ROC curve that is not convex
implies that a human observer, such as a 
radiologist, will give high ratings to
some of the images with signal or disease, but 
also give such images very low
ratings, even lower than the ratings given to
completely images with no signal.  The higher ratings are
expected, but the lower ratings are irrational.
There is no reason in general that a human observer 
would give the lowest of his
ratings to images that really contain a signal. 
Logically, scientists  believe that in the limit of
barely perceptible signals the worst that human observers could do
is guess, in other words, assign a low rating to 
a signal present or signal absent image with equal probability.  
Theoretically the likelihood ratio
of a pair of distributions, $\text{d}F/\text{d}H$, that model human detection
responses should be a monotonically increasing function
of the rating given by the observer.  Models of ROC data 
that have monotonic likelihood ratios are called 
``proper.''~\citep{GreenNSwets,dorfman-bigamma-1996,MetzProproc1999,lloyd2002}
\citet{Pesce2010} detail why we should expect to see convex
ROC curves under the assumption that radiologists would
perform their task at least as accurately as guessing, and  
they propose a method for constructing a convex ROC curve estimate.

In this paper we examine the nonparametric maximum likelihood estimate 
(NPMLE) of the ROC curve under the constraint that its likelihood ratio be
monotonic that was first developed by \citet{lloyd2002}.
Like \citet{LimWon} we prove that this estimate is 
equivalent to a simple convex hull around the usual empirical
or unconstrained nonparametric ROC curve MLE.  Unlike \citet{LimWon} 
we prove this directly from the likelihood and constraints thereon.

This paper examines the
properties of the constrained ROC NPMLE under a large number
of population parameters.  In particular we evaluate the 
properties of the associated AUC estimate, the area under the 
the constrained ROC NPMLE.  We demonstrate that this estimate
has less variance than an unconstrained estimator,
and it may have less bias for discretized data.

Unlike \citet{lloyd2002} and \citet{LimWon}, who
used numerical bootstrap techniques to estimate the variance of 
parameters of the constrained ROC curve MLE, we have found that
standard estimators of AUC variance may be used to obtain
variance estimates of constrained nonparametric 
maximum likelihood AUC estimate.
Over a wide range of simulation configurations we have found that
these estimates are nearly unbiased and very robust.
In summary this constrained estimator has a rational convex
ROC curve, is simple to construct, 
it may have small bias, its variance is easily estimated, and 
therefore it may be useful in observer performance experiments.

\section{Data}
\label{Data}

\subsection{Notation}

     \begin{figure}
\begin{center}
  \includegraphics[width=\columnwidth]{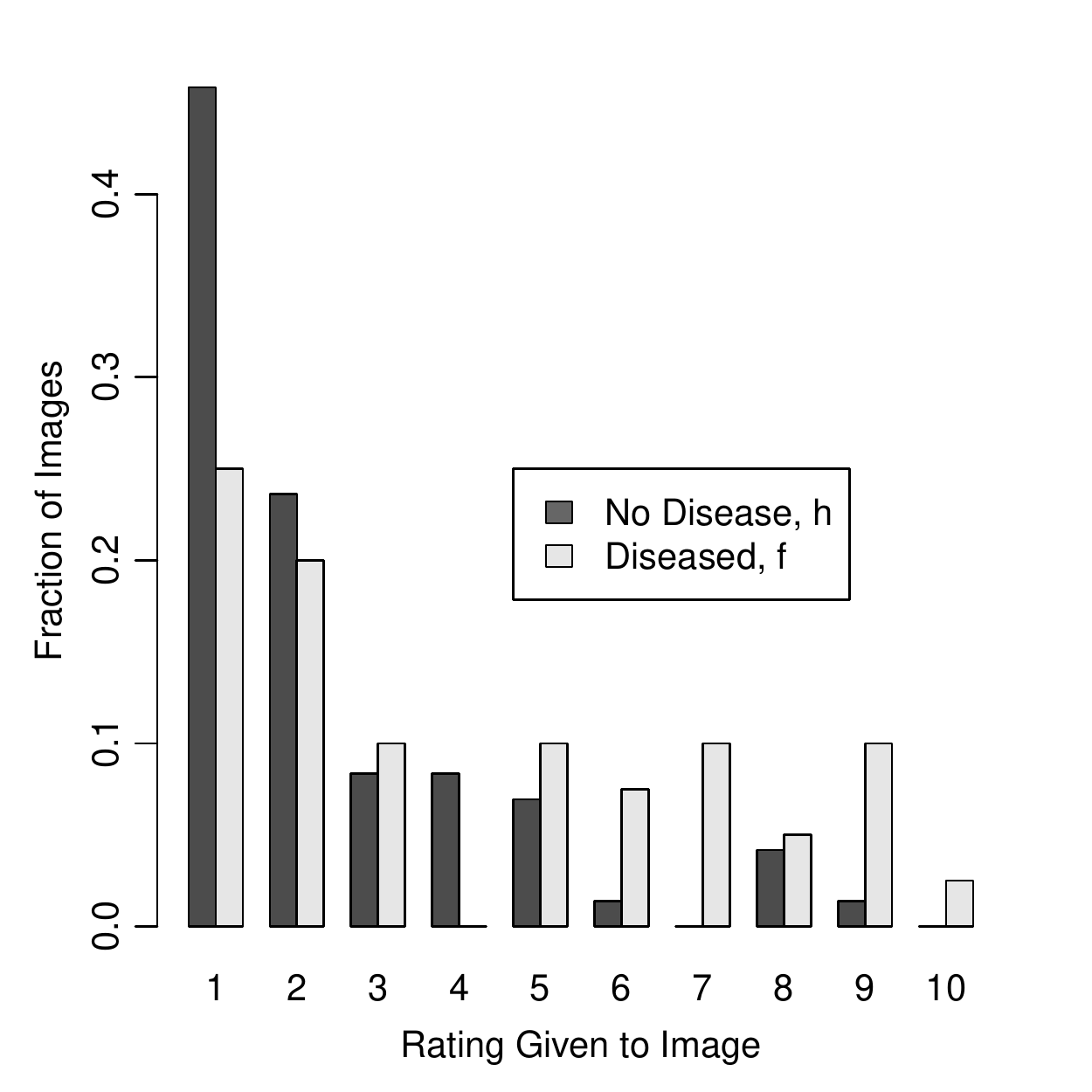}
       \caption{ This figure shows two discrete density 
         functions of ordinal observations
from two different classes of objects, images with disease
and images with no disease.  This data is from \citet{Chan2001} and
is also given in Table~\ref{reader5counttable}.  The empirical cumulative 
distributions of these data sets can be plotted against each other 
to form the ROC curve shown in Figure~\ref{Chan}.}

          \label{histogram}
          \end{center}
       \end{figure}

Our data of interest is typically ordered categorical data
collected from humans observing two classes of objects.

Let 

$ x_{1}, \dots, x_{N}$ 
be an IID random sample of ratings or scores from the signal-absent or non-diseased population
with density $h$, let 

$ y_{1}, \ldots, y_{M}$ 
be a random sample 
from the signal-present or diseased population with density $f$. All ratings are assumed to be discrete in
$k$ ordered categories. Each signal-present  score $y_s,~ s=1\dots M$ corresponds  to a category index $d_s$, and  each signal-absent score $x_r,~ r=1\dots N$ corresponds to a category $c_r$.  There are $m_i$ signal observations and $n_i$ signal-absent observations in the $i^{th}$ category.  For simplicity we assumed the scores in each signal-absent category $i$ to be all identically equal to $\bar{x}_i$ and the scores in each signal-present category $j$ to be all identically equal to $\bar{y}_j$. 

Let $N=\displaystyle\sum_{i=1}^{k}n_i$,  and  $M= \displaystyle\sum_{i=1}^{k}m_i$.

\subsection{Example Data Sets}
\label{dataset}
As examples 
we have selected data sets from a study by \citet{Chan2001}.
In the study, radiologists reviewed
X-ray images of mammographic microcalcifications (lesions)
that were digitized at different resolutions.
Every radiologist rated each lesion from one to ten 
at every resolution.
Higher ratings indicated a radiologist's greater confidence
in the malignancy of the lesion.  Radiologists gave
lower ratings to lesions that they believed to be benign.
The study measured the ability  of radiologists 
to differentiate malignant lesions from benign lesions
as a function of the image resolution 
as measured by the area under the ROC curve (AUC).
In the study the authors used a bi-normal parametric 
method as in \citet{dorfman-alf-1969} to estimate the ROC curve
from the ten-category data.

The distribution of scores from radiologist~5 
evaluating the lesions at a resolution 140 microns 
is given in Figure~\ref{histogram}.

Figure~\ref{Chan unconstrained}  shows the empirical ROC curve derived from
this data.  Easily visible in the figure are the 10 solid 
line segments derived from the 10 rating categories.
We observe that the empirical ROC curve is not convex for this given data set.
Later in Section~\ref{application} we use this data and 
ratings from radiologist~4 at a 
resolution of 35 microns to demonstrate the methods
described in this paper.

In this experiment we have confidence that  doctors will not systematically give lower
ratings to images that contain lesions, nor will they

systematically give higher ratings to images with no lesions.
Therefore we believe the true
underlying population ROC curves will be convex,
and any nonconvexity in these empirical ROC curve estimates
is merely due to statistical uncertainty from finite sample size.
The average ROC curve across all radiologists and experiments in 
Figure~7 of \citet{Chan2001} supports this assertion.

\section{Likelihood Function}
The usual  likelihood \citep{Owen1,Owen2} of obtaining the categorized ratings for 
both signal present and absent observations from our experiments
given our estimates of $F$ and $H$ is  
  \begin{equation}
  \label{E1}
\mathscr{L} \propto \displaystyle\prod_{i=1}^{k}p_i^{m_i}q_i^{n_i}
\end{equation}
where   $p_i= dF(\bar{y}_i)=F(\bar{y}_i)-F(\bar{y}_i-); q_i=dH(\bar{x}_i)=H(\bar{x}_i)-H(\bar{x}_i-)$, and $\bar{y}_i-$ is the largest possible value
of $y$ less than $\bar{y}_i$.
As in the power biased model  \citet{Qin94}, \citet{Vardi85}, and \citet{Vardi82}, we set  $w= p/q$.
The likelihood function becomes 
\begin{equation}
\label{E2}
 \mathscr{L} \propto \displaystyle\prod_{i=1}^{k}q_i^{n_i+m_i}w_i^{m_i}. 
 \end{equation}
  We use the Lagrange multipliers as in \citet{Anderson_72} to maximize the likelihood function.
Maximizing  $ \mathscr{L}$ subject to 
   $\displaystyle\sum_{i=1}^{k}q_i =1$ and $\displaystyle\sum_{i=1}^{k}q_i w_i=1 $ is equivalent to minimizing   
   \begin{equation}
\label{loglike}         
  -\log(\mathscr{L})= \displaystyle\sum_{i=1}^{k}- (n_i+m_i)\log(q_i)-m_i\log(w_i)+ C
  \end{equation} 
subject to the same constraints, where $C$ is a constant.

By first minimizing $ -\log(\mathscr{L})$ as a function of $q_i\dots,q_k$ we obtain $ q_i=\frac{m_i+n_i}{N+Mw_i}$. 
Plugging $\frac{m_i+n_i}{N+Mw_i} \text{ for } q_i$ in Equation \ref{loglike} and ignoring the constant terms we are  then led to minimize \[ \ell =\displaystyle \sum_{i=1}^{k}(m_i+n_i)\log(N+Mw_i)-m_i\log(w_i)\] as a function of $W =(w_1\dots,w_k)$.
   
 \section{Estimates}
 \subsection{Unconstrained Estimates} 
\label{unconstrained}
As a function of $ w_1,\dots, w_k$,  $\ell$ is minimized for $\bar{w}_i=\frac{m_i/M}{n_i/N}.$
We plug $\bar{w}_i$ back to $ q_i=\frac{m_i+n_i}{N+Mw_i}$ and $ p_i =q_iw_i $ to obtain the known empirical likelihood estimates
$ \bar{q}_i= \frac{n_i}{N} $ and $\bar{p}_i= \frac{m_i}{M}$. 

In what follows we will refer to $ \bar{w}_i=\frac{m_i/M}{n_i/N}$ as the unconstrained maximum likelihood ratio (UMLR).
We should point out that these estimates of $ \bar{q}_i$, $\bar{p}_i$, and  $ \bar{w}_i$  
give the usual empirical ROC curve~\citep{GreenNSwets,Bamber}.
The empirical likelihood ratios $w_i$ are the slopes of the 
segments of the lines of the empirical ROC curve, i.e. the slopes 
of the lines in Figure~\ref{Chan unconstrained}.

\subsection{Constrained Estimates}
\label{constrained}
Believing that the true ROC curve is convex,  we are led to minimize
 $\ell =\displaystyle \sum_{i=1}^{k}(m_i+n_i)\log(N+Mw_i)-m_i\log(w_i)$  subject to the monotonic likelihood ratio constraint 
  $w_1\le w_2\le \dots\le w_k$.
This is equivalent to requiring that the slopes of the segments of the 
ROC curve be increasing from right to left.
Let $ \bar{p}_i ~\text{and } \bar{q}_i$ be defined as in Section~\ref{unconstrained}.
 Let  $  P_0=Q_0=1$, and $P_i=\displaystyle\sum_{j=1}^{k-i}\bar{p}_j,~ Q_i=\displaystyle\sum_{j=1}^{k-i}\bar{q}_j$, for $ i=1\dots k$.

\begin{figure*}
  \centering
 \subfloat[Likelihoods of a Convex Estimate]{\label{likelihood convex}\includegraphics[scale=.40]{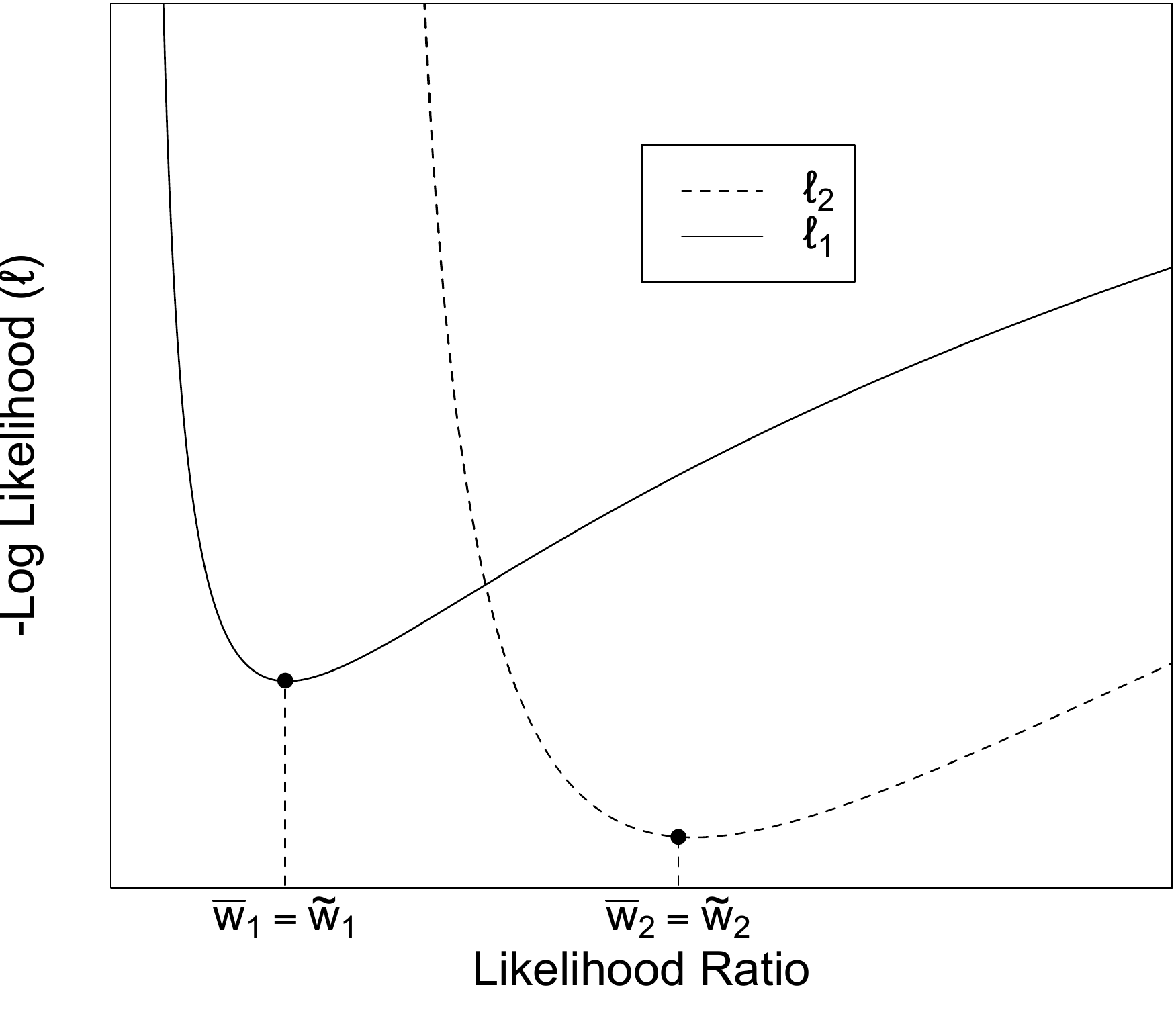}}
  \subfloat[Likelihoods of a Non-convex Estimate]{\label{likelihood nonconvex}\includegraphics[scale=.40]{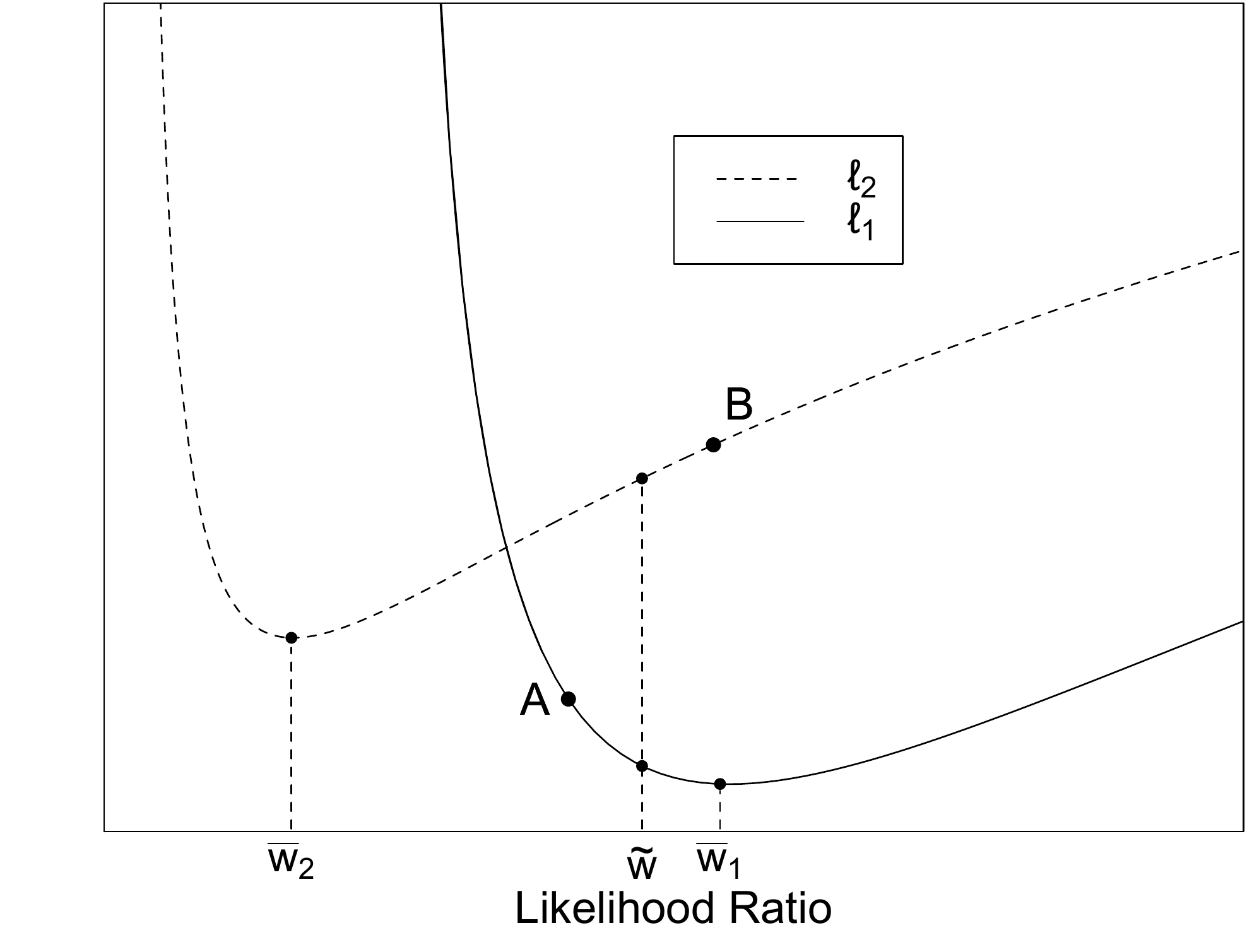}}
  \caption{These are graphs of the negative log likelihood components $\ell$ as a function of the likelihood ratios $w$.The unconstrained minimum of $\ell_i \text{ is } \bar{w}_i$. In Figure~\ref{likelihood convex} the unconstrained minimum $\bar{w}_1$ of the $-\log$ likelihood function $\ell_1$  is less than  the unconstrained minimum $\bar{w}_2$  of   $\ell_2$ ; therefore, the likelihood function $\ell=\ell_1+\ell_2$  yields a convex ROC. In Figure~\ref{likelihood nonconvex} $\bar{w}_1>\bar{w}_2$ meaning the corresponding -$\log$ likelihood ratio yields a non-convex ROC curve.}
  \label{likelihood}
               \end{figure*}

    As a function  of $ w_1,\dots,w_k,$   $\ell$ is a sum of k independent functions  $ \ell_1,\dots, \ell_k$;  where
  $\ell_i=(m_i+n_i)\log(N+Mw_i)-m_i\log(w_i)$ is a function with a single minimum   $\bar{w}_i$.  As shown in Figure~\ref{likelihood}, $\ell_i$ decreases up to $\bar{w}_i$ and increases afterward.

Without loss of generality we consider two consecutive functions $\ell_1(w)$ and $\ell_2(w)$.
In Figure~\ref{likelihood} we plot $\ell_1\text{ and } \ell_2$ as  functions of $w_1\text{ and } w_2$ respectively.
 The graph in Figure~\ref{likelihood convex} shows the UMLR $\bar{w}_1 \text{ of } \ell_1 \text{ less than the UMLR } \bar{w}_2\text{ of } \ell_2$,  leading to a convex ROC curve. 
   On the contrary, in Figure~\ref{likelihood nonconvex} we have $ \bar{w}_1>\bar{w}_2$ leading to a nonconvex ROC.

  Let $ \bar{w}_1>\bar{w}_2$ as in  Figure~\ref{likelihood nonconvex}. 
   Our goal is to find $(\tilde{w}_1,\tilde{w}_2)$ such that 
 $\ell_1(\tilde{w}_1)+\ell_2(\tilde{w}_2)  =\min\limits_{w_1\leq w_2} \ell_1(w_1)+\ell_2(w_2)$. In this case, we need to change the ordering of 
 $w_1$ and $w_2$ while keeping $\ell_1(w_1)+\ell_2(w_2)$ as small as possible.\\\
 
 Let  $A(w_1,\ell_1(w_1))\text{  and } B(w_2,\ell_2(w_2))$ be two arbitrary points as in Figure~\ref{likelihood nonconvex} such that  $w_1< w_2$.
 Keeping in mind that $ \bar{w}_1$ and $\bar{w}_2$ are the respective global minima of $ \ell_1 \text{ and } \ell_2$ we obtain the following inequalities.
    \begin{eqnarray*} 
\text{ If } w_1<\bar{w}_2&\text{ then  }& \ell_1(\bar{w}_2)+\ell_2(\bar{w}_2) <\ell_1(w_1)+\ell_2(w_2)\label{eq1}\\
\text{ If } \bar{w}_1<w_2&\text{then}& \ell_1(\bar{w}_1)+\ell_2(\bar{w}_1) <\ell_1(w_1)+\ell_2(w_2) \label{eq2}\\  
\end{eqnarray*} 

 If $\bar{w}_2<w_1<w_2<\bar{w}_1$, $~\ell_1$ decreases while $\ell_2$ increases on the interval $[\bar{w}_1,\bar{w}_2]$, leading to the following inequalities.
 
      \begin{eqnarray*} 
\ell_2(w_1)<\ell_2(w_2)&\Rightarrow&\ell_1(w_1)+ \ell_2(w_1)~<~\ell_1(w_1)+\ell_2(w_2) \\
     &\text{ and }& \\
\ell_1(w_2)<\ell_1(w_1)& \Rightarrow & \ell_1(w_2)+\ell_2(w_2)~<~ \ell_1(w_1) +\ell_2(w_2).\\
    \end{eqnarray*}
    The above inequalities show that the function $\ell_1+\ell_2$ is never at its minimum for two distinct values of $w_1 \text{ and } w_2$.
     We deduce that $(\tilde{w}_1,\tilde{w}_2)$ satisfies  $\tilde{w}_1=\tilde{w}_2.$ 
  Setting $ w_1=w_2=w$ we minimize the following expression:
    \begin{eqnarray*}
    \ell_1(w)+\ell_2(w) &=&(m_1+n_1)\log(N+Mw)-m_1\log(w) + \\
                       &  & (m_2+n_2)\log(N+Mw)-m_2\log(w)\\
    &=& (m_1+m_2+n_1+n_2)\log(N+Mw)- \\
    & & (m_1+m_2)\log(w).\\
    \end{eqnarray*}
    Taking the derivative and setting equal to zero we obtain 
    $\tilde{w}=\tilde{w}_1=\tilde{w}_2= \frac{(m_1+m_2)N}{(n_1+n_2)M}$.
    Note that for $ w_1=w_2$ we obtain $ \ell^{\ast}(w)=\ell_1(w)+\ell_2(w)$ whose graph is similar to that of either $ \ell_1 \text{ or } \ell_2 $ from 
    Figure~\ref{likelihood}. 
The above proof and conclusions apply to any two adjacent categories
in a ROC curve where the unconstrained likelihood ratios are not increasing.

The above demonstrates analytically that the obtained estimate is the constrained maximum likelihood estimate of the non-parametric model. 

Note that the constrained estimators 
$\tilde{w}=\tilde{w}_1=\tilde{w}_2$ yield a solution that 
is equivalent to adding the two adjacent categories in to a single
category with counts $\tilde{m} = m_1 + m_2$ and 
$\tilde{n} = n_1 + n_2$. 
To find all constrained estimates $\tilde{w}_i$ in a sequence of 
$\ell_i$, we use an algorithm similar to \citet{lloyd2002}.
    
            \section{PAVA: Pool Adjacent Violator Algorithm }
            \label{pava}
          The PAVA, a simple iterative algorithm,  is a standard nonparametric method used to produce point estimates for a function known to be monotone \citep{Barlow1972,Dykstra,Chakravarti}.  

It replaces each stretch of monotonicity-violating observations with the weighted average of the original function values in that stretch.
            
           In order to apply  the theoretical approached described above, we developed an R \citep{Rcite} version of the  PAVA  algoritm similar to that of \citet{lloyd2002}. 
     As in Section~\ref{Data}, we assume the data has been discretized into k different categories. 
     Keeping in mind that each signal-present  score $ y_s$  corresponds to a category $d_s$ and each signal-absent score $x_r$  corresponds to a category $c_r$  as assumed in Section~\ref{Data}, as the PAVA combines categories it also reassigns each data score to a new category.

     We  sequentially describe the PAVA as follows.
        \begin{enumerate}
         \item $ \tilde{k} \leftarrow k;~$ for all $i$, $\tilde{w}_i \leftarrow \bar{w}_i=\frac{m_iN}{n_iM}; \tilde{m}_i \leftarrow m_i; \tilde{n}_i \leftarrow n_i$; for all $s$, and for all $r$, $\tilde{d}_s \leftarrow d_s;~\tilde{c}_r \leftarrow c_r$ 
         \item Stop if ROC curve is convex; meaning, $\tilde{w}_1\le...\le \tilde{w}_{\tilde{k}}$.  Otherwise
 there exists $i$ in $  1:\tilde{k}-1$ such that $\tilde{w}_i >\tilde{w}_{i+1}.$ If so, continue.          
     \item  The two  categories $i$ and $i+1$ are combined into category $i$.      $\tilde{m}_i\leftarrow \tilde{m}_i+\tilde{m}_{i+1},~ \tilde{n}_i\leftarrow \tilde{n}_i+\tilde{n}_{i+1}; \tilde{w}_i\leftarrow \frac{\tilde{m}_iN}{\tilde{n}_iM}$.

    \item  After combining the two categories $i$ and $i+1$  into category $i$, the following commands describe how we eliminate category $i+1$, reorganize the remaining categories, and reassign the score category indexes .\\ 
    For $ s =1\dots M, \text{ if } \tilde{d}_s>i\,  \text{ then } \tilde{d}_s\leftarrow \tilde{d}_s-1;$  \\
    for $ r = 1\dots N, \text{ if }\tilde{c}_r>i   \text{ then }\tilde{c}_r\leftarrow \tilde{c}_r-1;$\\
    for $j=i+1\dots \tilde{k}-1,~ \tilde{n}_j \leftarrow \tilde{n}_{j+1};~\tilde{m}_j \leftarrow \tilde{m}_{j+1};~\tilde{w}_j \leftarrow \tilde{w}_{j+1}$;  \\
    $ \tilde{k} \leftarrow \tilde{k}-1.$

       \item \vspace{.14in} Go to step 2. Proceed until ROC curve is convex.  
        \end{enumerate}
        Applied to the data, the PAVA  creates a new data set with total number of categories $\tilde{k}$ which is less than or equal to $k$.  We respectively denote the number of signal-absent scores in each new i-category, the number of signal-present scores in each new  i-category, the new signal-present category index, and the new signal-absent category index of the new data set by $\tilde{n}_i$, $\tilde{m}_i$, $\tilde{d}_s$, and $\tilde{c}_r$.

\begin{figure*}
     
   \centering

   \subfloat[NPMLE ROC Curve]{ \includegraphics[scale=.40]{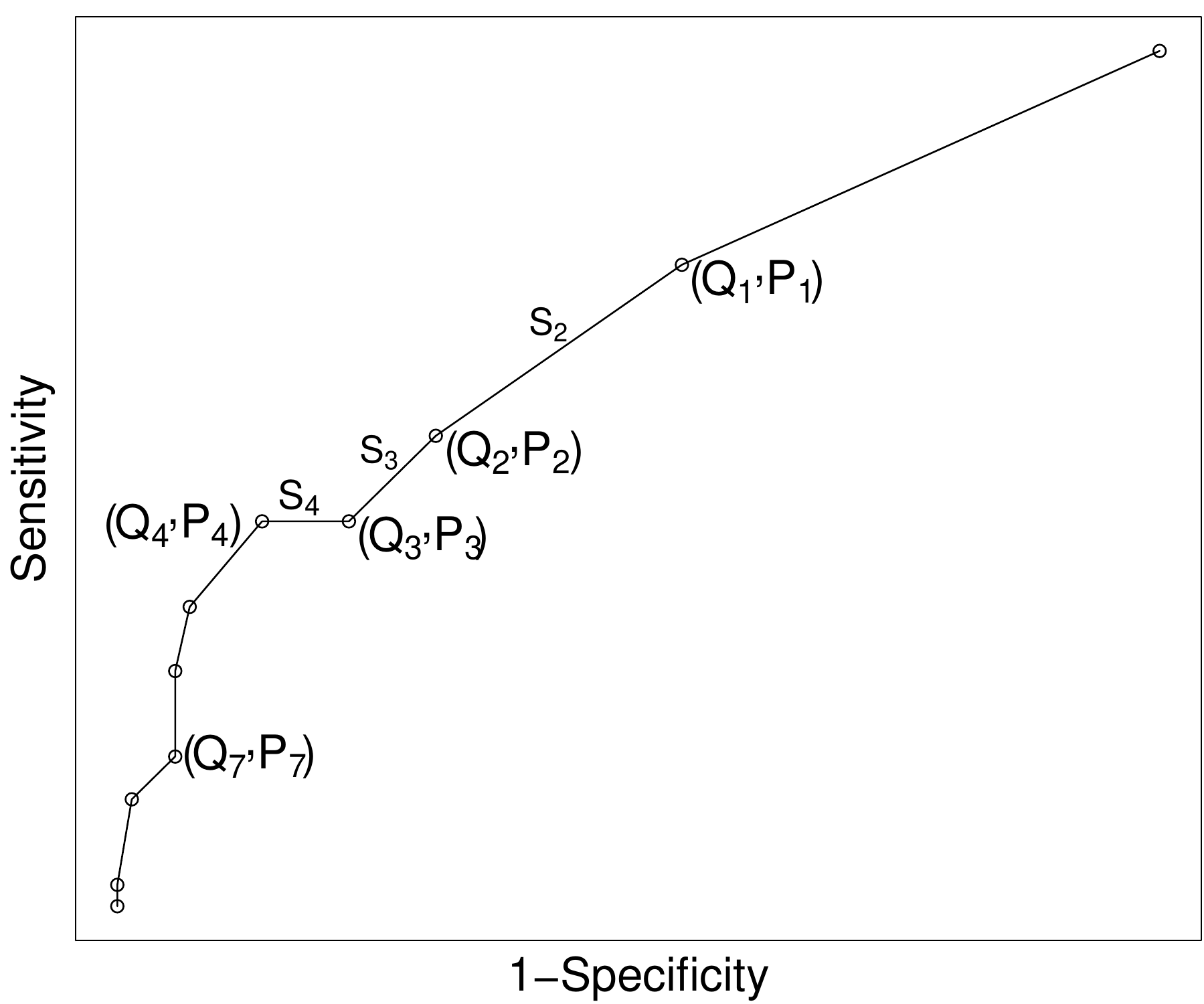}\label{Chan unconstrained}} 

  \subfloat[Constrained NPMLE ROC Curve]{\label{Chan constrained} \includegraphics[scale=.40]{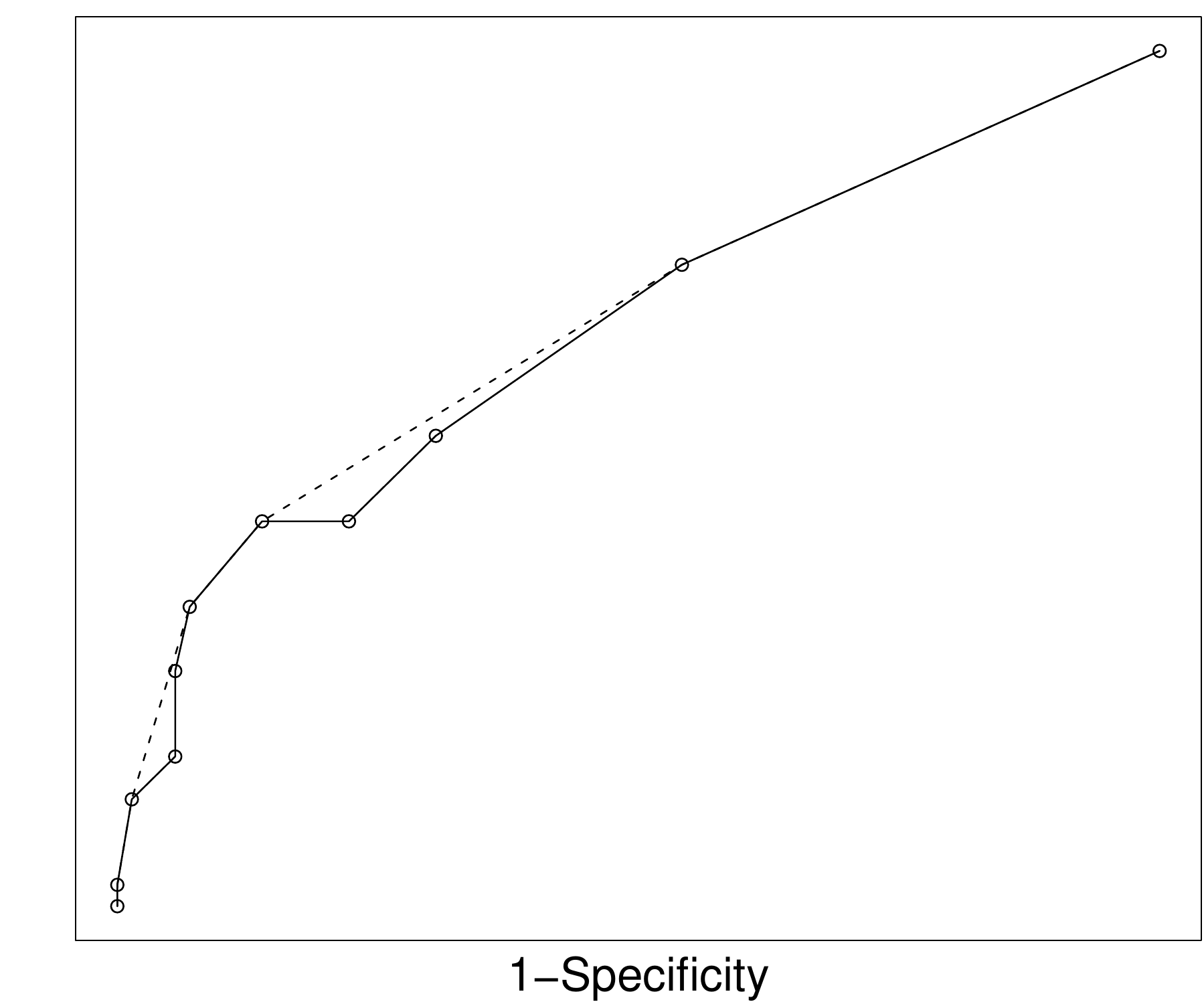}}
 \caption{ \label{Chan} In Figure~\ref{Chan unconstrained} we plotted the ROC curve of the data from \citet{Chan2001} 
described in Section~\ref{dataset} and given in Table~\ref{reader5counttable}.
The ROC curve is the set all possible measurable specificity and sensitivity pairs.
We obtained the constrained estimate (dotted line) in Figure~\ref{Chan constrained} by 
applying the PAVA to the data used in Figure~\ref{Chan unconstrained}.
In Figure~\ref{Chan unconstrained} we have $k=10$ line segments. $P_i \text{ and}~ Q_i$ are defined in Section \ref{constrained}.
Segments $S_2$, $S_3$, and $S_4$ form a violation because the slope $\bar{w}_4$
of segment $S_4$ is so low.  Therefore we replace those 3 segments with a
single segment $ S^\prime_2$ with slope  $\tilde{w}_2={(m_2+m_3+m_4)N}/{(n_2+n_3+n_4)M}$   Also see Table~\ref{reader5counttable}.
Categories $S_6$, $S_7$, and $S_8$ also form a violation and are replaced 
with a single category.}

 \end{figure*}

The dashed line in Figure~\ref{Chan constrained} is obtained from applying the PAVA to the ROC curve in Figure~\ref{Chan unconstrained}.
In Figure~\ref{Chan unconstrained} we have k=10. Let $p_i,~q_i,~P_i \text{ and}~ Q_i$ be defined as in Section \ref{constrained}.
In the first step there is a violation for $i=3$, i.e $\bar{w}_3$  is greater than $\bar{w}_4$. The  two segments 
 $ S_3 \text{ and }  S_4$ 
 are replaced by a single segment $S^\prime_3$ from $ (Q_2,P_2) \text{ to }(Q_4,P_4)$ and $\tilde{w}_3 $ is set set equal to
$ (\frac{m_3+m_4}{M}/\frac{n_3+n_4}{N})$. This leads to another violation as the new $ \tilde{w}_3 $ is less than $\bar{w}_2$.  We then replace $S_2 \text{ and } S^\prime_3 $ by  a single segment $ S^\prime_2 $ from  $(Q_1,P_1) \text{ to } (Q_4,P_4) \text{ and } \tilde{w}_2$ is set equal to $\frac{(m_2+m_3+m_4)N}{(n_2+n_3+n_4)M}$.

In the next step the PAVA encounters another violation between the segments  $S_6 \text{ and }  S_7 $; the segments 
 $ S_6 \text{ and } S_7$ are replaced by a single segment $S^\prime_6$ . 
  In the final step, the PAVA encounters another  violation as the slope of $S^\prime_6$  is greater than the slope of $S_8$; we then replace $S\prime_6 \text{ and } S_8$ by a single segment and the ROC is convex.

  An intuitive reason for implementing the algorithm can be 
summarized as this:

Given that observers will not perform worse than guessing
we assume that any non-convexity in a
measured empirical ROC curve
is due to a statistical variation from insufficient counts
in a particular rating category.  We therefore combine 
adjacent ordinal categories until the number of counts 
in the combined categories are sufficient to generate a 
plausible convex ROC curve.

   This PAVA algorithm is equivalent to the algorithm given in Section 1.2 of
   \citet{Dykstra}.  Specifically using Robertson's 
   w, $g$, $W$, and $G$ notation
   let  w$(\bar{x}_i)=\frac{n_i}{N},~~g(\bar{x}_i)=\frac{m_i}{M}/\frac{n_i}{N},~~W_j=\displaystyle\sum_{i=1}^j\text{w}_j,~ G_j=\displaystyle\sum_{i=1}^jg(\bar{x}_i)\text{w}(\bar{x}_i)$.
   Then the ROC curve, a plot of  $ P_j=(W_j,G_j),~ j=0,1,\dots,k \text{ with } P_0=(0,0)$ is the \emph{cumulative sum diagram} (CSD) for the function g with weight w.   
  Adapting Robertson's approach,  finding the best convex plot from the nonconvex one reduces to finding the 
    \emph{ greatest convex minorant} (GCM) of the CSD in the interval $[0,Z_k]$  for the function g with weight z.
     This method will lead to the same estimate obtained in Section~\ref{constrained}.
       
  To verify that the above solution really is the maximum likelihood
  estimate under the constraint of a monotonic likelihood ratio,
  we wrote a software function whose input is a combined vector of $ p_i
  \text{ and }q_i $ and whose output is the likelihood
  function given in Expression~\ref{E1} if the ROC curve is convex. 
  If the ROC curve is not convex, the output is a value
  less than zero. 
  We then used a numerical optimizer 
  (``optim'' from the package R~\citep{Rcite}) 
  to maximize the function over all inputs $p_i, q_i$ on many random
  simulated data sets with random initial values.  The maximum
  likelihood function given by the optimizer was usually equal to that
  obtained from the PAVA.  All other times the optimizer failed to
  converge or converged to a local maximum with likelihood
  less than that obtained from the PAVA.  This verified numerically
  that the PAVA gives the constrained maximum likelihood.

\section{Bias and Variance Estimation of AUC} 
\label{Variance Estimation}
The Wilcoxon-Mann-Whitney statistic is equivalent to the 
area under the ROC curve (AUC) and is frequently used as a summary 
measure of separability of the observations that contain a signal from
observations that do not, or separability of images that have disease
from those who do not.  The area under the population ROC curve is equal
to the probability that a 
rating of an image with a signal is higher than a 
signal-absent rating, AUC$=P(Y>X)$~\citep{Bamber}.  
The area under the unconstrained empirical
maximum-likelihood ROC curve estimate is
\begin{equation}
\label{aucsum1}
\widehat{AUC} = \frac{1}{NM}\sum_{r=1}^N\sum_{s=1}^M I_{rs}
\end{equation}
We assume that each pair (r,s) corresponds to a pair $(c_r,d_s)$  such  that $x_r$ belongs to the category $c_r$ and $y_s$ belongs to the category $d_s$.
\begin{equation}
\label{auc}
I_{rs}=\left\{
\begin{array}{rlll}
 \frac{1}{2} &\text{ if } c_r=d_s\\
0& \text{ if } c_r>d_s\\
1& \text{ if } c_r< d_s
\end{array} 
\right.
\end{equation}

The area under the constrained empirical
maximum-likelihood ROC curve estimate has the same form and is
\begin{equation}
\label{aucsum2}
\widetilde{AUC} = \frac{1}{NM}\sum_{r=1}^N\sum_{s=1}^M I^\prime_{rs} 
\end{equation}
  where \begin{equation}
\label{aucc}
I^\prime_{rs}=\left\{
\begin{array}{rlll}
 \frac{1}{2} &\text{ if } \tilde{c}_r=\tilde{d}_s\\
0& \text{ if } \tilde{c}_r>\tilde{d}_s\\
1& \text{ if } \tilde{c}_r< \tilde{d}_s
\end{array} 
\right.
\end{equation}
 
In this Section we use simulated data to measure how the constrained AUC estimate
differs from the unconstrained AUC estimate and a true continuous
AUC population value.  We also examine ways to estimate variance of the constrained AUC estimate.   
\subsection{Simulation Parameters}
\label{simpar}
We simulated data from normal distributions, D=$\mathcal{N}(\mu,1)$,  and  from uniform distributions, D=$\mathcal{U}(0,\mathfrak{m})$, which
are uniform in probability between 0 and $\mathfrak{m}$.
For the uniform distributions $\mathcal{U}$, the non-diseased 
or signal-absent sample 
is drawn from $\mathcal{U}(0,1)$, and the signal-present or diseased sample is drawn from
$\mathcal{U}(0,\mathfrak{m})$ where $\mathfrak{m}\ge1$.   Both samples together form a combined 
sample.
In order to have the desired $AUC$  value for the uniform distribution, it follows from $AUC = P(Y>X)$   that  $AUC=1-\frac{1}{2\mathfrak{m}}$  from which we derive $\mathfrak{m}=\frac{1}{2(1-AUC)}.$
This pair of uniform distributions yields the most asymmetric ROC 
curve that still has a monotonically increasing likelihood ratio.

For the normal distribution $\mathcal{N}$, the diseased sample is drawn from  $N(\mu,1)$ and the non-diseased sample is drawn from $\mathcal{N}(0,1)$. 
To achieve the desired $AUC$, we used   $P(Y>X)= \Phi\left( \frac{\mu_Y-\mu_X}{\sqrt{\sigma_X^2 +\sigma_Y^2}}\right)$,
where $\Phi$ is the cumulative normal distribution function, $\mu_Y=\mu$, and $\mu_X=0$.  We set  $ \mu=\sqrt{2}\Phi^{-1}(AUC) $.   This model 
gives a continous symmetric convex ROC curve.

\subsection{Bias Study}
\label{bias}
We first considered a vector of AUC values with elements ranging from .6 to .99 with step equal to .01.
For each AUC value from the above vector, we sampled data from either normal distributions, D=$\mathcal{N}$, or uniform distributions, D=$\mathcal{U}$. 
We simulated 10,000 independent random combined samples from D.  
Each combined random sample consisted of a sample of size 55  drawn 
from the non-diseased distribution and a sample of size 110 drawn from
the non-diseased population. These sample size are typical in many studies in medical literature.
We then discretize each sample into seven categories. For each discretized combined sample we computed the usual unconstrained AUC estimate, then applied the PAVA algorithm and computed the constrained AUC estimate.

\begin{figure}
  \caption{
    This figure shows the fractional bias (the mean estimated
value divided by the true value) of the two nonparametric 
   estimators of AUC in this paper as a function of AUC.   
The mean of the
usual unconstrained nonparametric AUC estimate (equation~\ref{aucsum1}) is always less
than the true underlying simulation AUC value because the observed values
are discretized.
The constrained AUC estimate (equation~\ref{aucsum2})
is always greater than the unconstrained AUC estimate.
The constrained AUC estimate can be considered 
less biased  than the usual unconstrained emprirical AUC estimate for most values of the AUC.}
  \label{cat7}
  \centering
  \includegraphics[scale=.40]{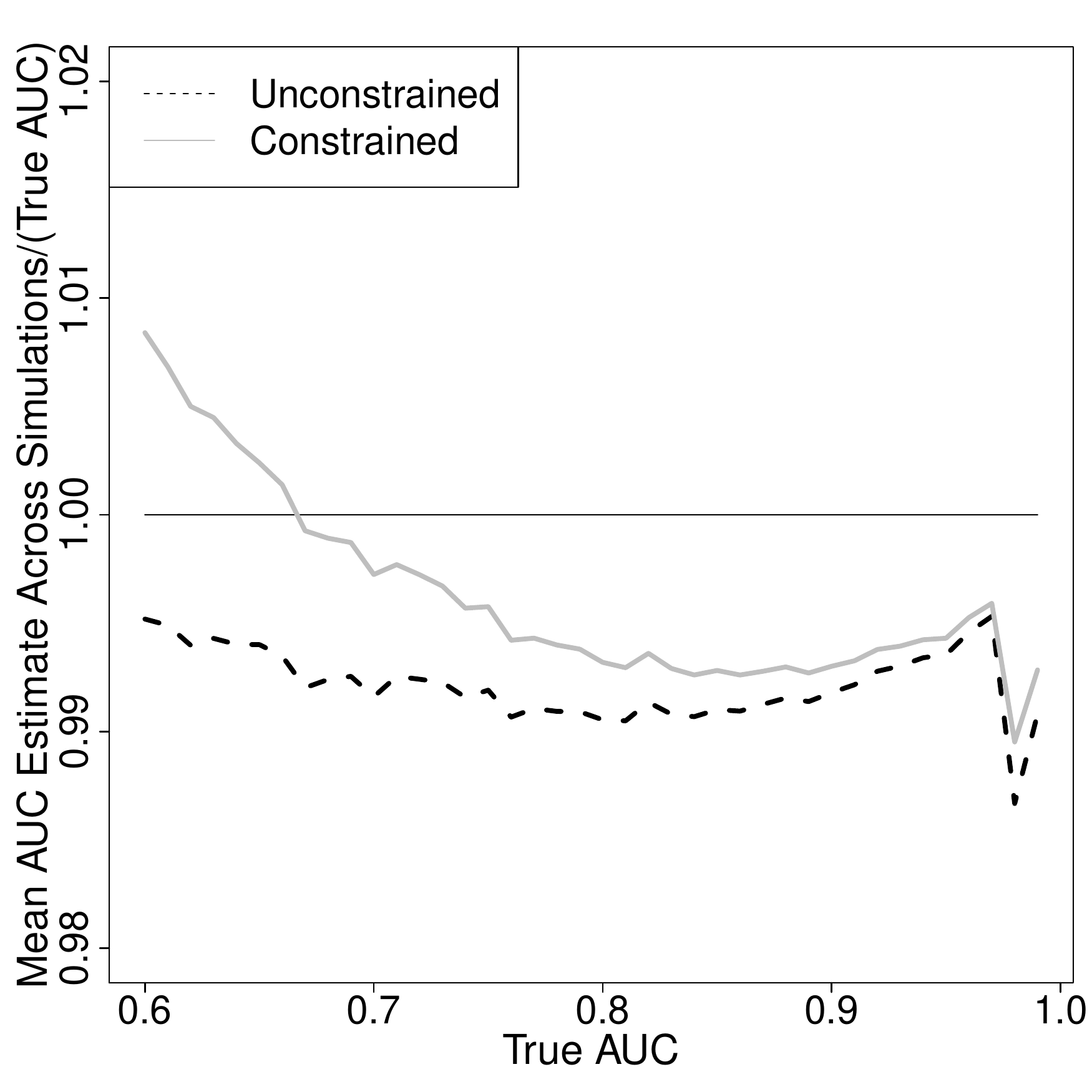}
               \end{figure}

In Figure \ref{cat7} we plotted the average constrained and unconstrained
AUC estimates from the normally distributed simulation versus the true AUC.
We first note that the average unconstrained AUC estimate is always 
biased low with respect to the true undiscretized population AUC.  
This reduction of
information and AUC estimates due to discretization is well known.
We also note that the average constrained AUC estimate is always greater 
than the average unconstrained AUC estimate.  This result is consistent
with the construction of the constrained ROC curve in 
Section~\ref{constrained}.  To construct the constrained ROC estimate 
we combine categories in a way that always increases the area under the 
curve.  With respect to the true population AUC, the
average constrained AUC estimate is sometimes less, sometimes greater, 
depending on the value of AUC.   Depending on AUC, sample size, 
and discreteness of data, the constrained estimate may be 
less biased than the unconstrained estimate.

\subsection{Variance Estimation of the AUC}

In this Section we demonstrate empirically 
that a standard variance estimator of
the empirical area under the curve (AUC) also provides a good estimate
the AUC of the above convexity-restricted ROC curve estimate.
Our method of variance estimation can be considered as a two-way analysis of variance (ANOVA) with one observation per cell as in Chapter~7 of \citet{Scheffe}. We model $I_{rs}$ where each cell  consists of~1 if the diseased score~$s$ is greater than the non-diseased score $r$, of $\frac{1}{2}$ if both scores are equal, and~0 otherwise. See equation~\ref{auc}.
The model is  $I_{rs}=\mu+a_r+b_s+\epsilon_{rs}$ for  $r=1...N$;$ s=1\dots M$,
or equivalently for AUC, $\hat{A}=\frac{1}{NM}\displaystyle\sum\limits_{r=1}^N \sum\limits_{s=1}^M (\mu+a_r+b_s+\varepsilon_{rs})$
 where $\mu$ represents the overall mean AUC.  The random variable $a_r$ has mean zero and variance $\sigma_a$ reflecting  the variability  between the disease tests. $b_s$ has mean 0 and variance $\sigma_b$ reflecting the variability between the non-disease tests. $\varepsilon_{rs}$ has mean zero and variance $\sigma_{\varepsilon}$ representing the interaction between disease and non-disease tests.
The total variance of $\hat{A}$ is 
\begin{equation}
\label{var}
\text{Var}(\hat A)=\frac{\sigma^2_a}{N}+\frac{\sigma^b_b}{M}+\frac{\sigma^2_\varepsilon}{NM}  
\end{equation}
The respective estimates 
$\hat{\sigma}_a,~ \hat{\sigma}_b,\text{ and } \hat{\sigma}_{\varepsilon}$ of
$\sigma_a,~ \sigma_b,\text{ and } \sigma_{\varepsilon}$
 are computed as in equation 7.4.10 of \citet{Scheffe}.
For simulation purposes, we chose the distribution D=$\mathcal{U}$ or D=$\mathcal{N}$ as in  Section~\ref{bias}
and then chose the parameters of the distribution so that AUC=.6, AUC= .84, and  AUC=.94 which respectively correspond to $\mathfrak{m}=1.25, \mathfrak{m}=3.125, \text{ and } \mathfrak{m}=25/3$ for D=$\mathcal{U}$ and to $\mu=.358,~\mu=1.406,~\text{ and } \mu=2.199$ for D=$\mathcal{N}$.
 There are often more non-disease patients available than disease patients. With that in mind, we assumed in the simulation  the ratio of non-disease  sample size to diseased sample size to take values $r_a=1, r_a=3$, and $r_a=9$.  This means for each  combined sample set consists of $M$ observations treated as diseased and $N=r_a\times M$ treated as non-diseased. We discretized each sample into k categories, where k takes values k=3, k=7, and k=16. For each quadruple $(D,M,r_a,k)$, we  generated 10,000 independent sets of combined random samples from the chosen distribution. 

For each quadruple $(D,M,r_a,k)$, we computed the unconstrained AUC $ \hat{A}_u^j$ and the estimate of variance 
$\widehat{V}_u^j$ using Expression~\ref{var}. We then  applied  the PAVA algorithm,  computed the constrained AUC $\widehat{A}_c^j$,  computed the estimate of the constrained AUC variance $\widehat{ V}_c^j$ using Expression~\ref{var}.  We then computed an approximate confidence interval of the AUC,
 $ I^{c j}_{1-\alpha}=\left[\widehat{A}_c^j-z_{\alpha/2}\sqrt{\widehat{V}_c^j}, \widehat{A}_c^j+z_{\alpha/2}\sqrt{\widehat{V}_c^j}\right], \text{ for }\alpha=.05$,
   where $z_{\alpha}$ is the standard normal deviate with probability $\alpha$.

\begin{figure}
  \caption{
This figure plots 
the square root of the mean of the variance estimates of AUC
as a function of 
the standard deviations of the AUC estimates of our simulations.
Data for both the usual unconstrained empirical AUC estimate and
the constrained AUC estimate are displayed.
Each point represents $10^4$ simulated samples from one
of our simulation configurations. 
The true standard deviation of AUC of the constrained ROC curves are lower on average than the standard deviation of the AUC of the unconstrained ROC curves, and the mean estimate of standard deviation is reduced by almost exactly the same amount.}
  \label{sd}
  \centering

 \includegraphics[scale=.50]{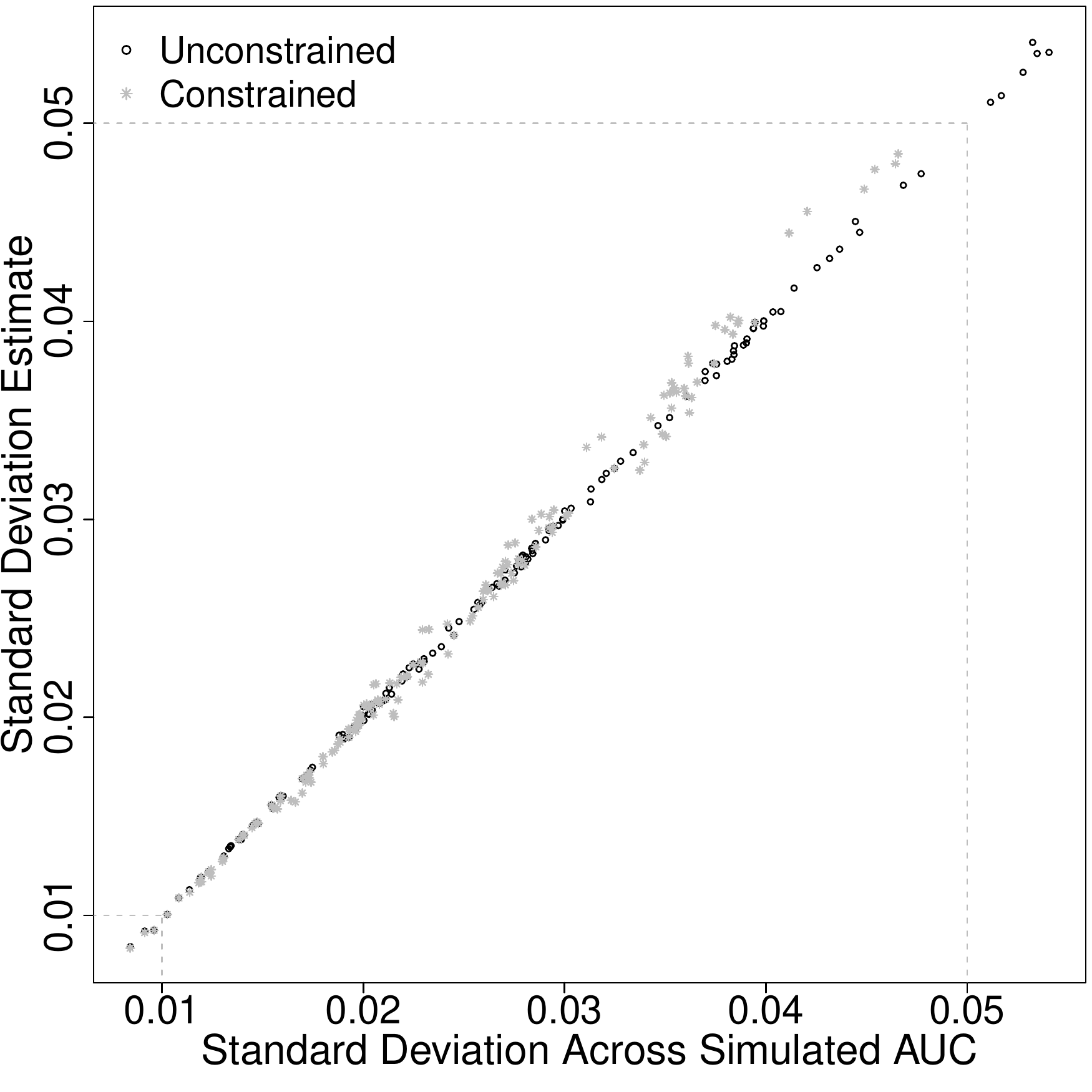}
               \end{figure}

 For each combined sample set of $10^4$ simulations, 
we plotted the square root of the mean of all the unconstrained variance estimates $ \widehat{V}_u^j$ versus the standard deviation across all the simulated unconstrained AUC $\widehat{A}_u^j$ in Figure~\ref{sd}.  
These data points lie on the diagonal of the plot indicating that 
our estimate of the variance of the empirical AUC estimate 
is unbiased \citep{Gallas}.

Also in Figure~\ref{sd} we plotted the square root of the mean of the constrained variance estimates $ \widehat{V}_c^j$  versus the standard deviation across all the simulated constrained AUC $\widehat{A}_c^j$. 
We note from these plots that the true standard deviation of AUC of the constrained ROC curves are lower on average than the standard deviation of the AUC of the unconstrained ROC curves, and the mean estimate of the standard deviation is reduced by almost exactly the same amount.  

From the figure it is apparent that our standard estimator of 
empirical AUC variance also very well estimates the variance 
of the constrained empirical estimate of AUC, and it is 
robust against the large range of parameters over which we
varied the simulation.

To test this assertion we report  in Table~\ref{cover} the coverage probabilities, which are the fractions of the simulations falling outside the above estimated confidence interval $I^{c j}_{1-\alpha} $. From Table~\ref{cover} we observe that the overall performance of the estimate of variance is accurate except in a few cases where the following conditions exist simultaneously: a large AUC, small sample size and a large number of categories.

Note that there is nothing special
about the AUC variance estimator that we
used.  Other variance estimates \citep{Bamber,Campbell} also
make useful measures the variance of the constrained AUC estimate.
These simple analytic variance estimators agree well 
with bootstrap estimation techniques \citep{EfronTibs,lloyd2002,LimWon}.
This includes the simple AUC variance approximation  
$\widehat{\text{AUC}}(1-\widehat{\text{AUC}})(N^{-1}+M^{-1})/4$
\citep{wagner_simplevar}.  
This form makes clear
the binomial-like behavior of AUC and the expected reduction in
variance corresponding to the increase in AUC.

\begin{table*}[h!]
\caption{Coverage Intervals with Nominal  Level 0.95}

  \begin{center}
\begin{tabular}{llllllllllcccrrr}

\multicolumn{14}{c}{UNIFORM DISTRIBUTION}\\
\hline
AUC&&&\multicolumn{3}{c}{3 categories}&&\multicolumn{3}{c}{7 categories}&&\multicolumn{3}{c}{16 categories}\\
&ratio $r_a$ && 1& 3&9&&1&3&9&&1&3&9\\
\hline
&&M=55&.0526&.0529& .0554&&.0710&.0671&.0577&&.0632&.0623&.0587 \\
.6&&M=100&.0493&.0509&.0536&&.0676&.0625&.0583&&.0577& .0542&.0578\\
&&M=200&.0627&.0572&.0562&&.0593&.0547&.0558&&.0534&.0508&.0553\\
\hline
&&M=55&.0639&.0644&.0657&&.0665&.0697&.0681&&.0754&.0736&.0796\\
.84&&M=100&.0597&.0540&.0581&&.0564&.0597&.0612&&.0644&.0625&.0605\\
&&M=200&.0493&.0537&.0574&&.0566&.0569&.0595&&.0552&.0574&.0580\\
\hline
&&M=55&.0678&.0717&.0696&&.0967&.1006&.1048&&.1056&.1158&.1206\\
.94&&M=100&.0623&.0659&.0649&&.0762&.0808&.0872&&.0902&.0901&.0954\\
&&M=200&.0559&.0569&.0557&&.0650&.0636&.0639&&.0718&.0672&.0684\\
\hline
\multicolumn{14}{c}{NORMAL DISTRIBUTION}\\
\hline
&&M=55&.0519&.0527&.0564&&.0628&.0576&.0575&&.0597&.0554&.0574&\\
.6&&M=100&.0522&.0568&.0523&&.0550&.0528&.0497&&.0531&.0483&.0490\\
&&N=200&.0489&.0519&.0502&&.0532&.0507&.0519&&.0520&.0474&.0468\\
\hline
&&M=55&.0586&.0585&.0627&&.0613&.0598&.0642&&.0722&.0643&.0663\\
.84&&M=100&.0514&.0603&.0560&&.0596&.0571&.0585&&.0602&.0564&.0522\\
&&M=200&.0548&.0548&.0513&&.0557&.0509&.0569&&.0610&.0564&.0537\\
\hline
&&M=55&.0725&.0780&.0830&&.0862&.0764&.0765&&.0955&.0846&.0891\\
.94&&M=100&.0610&.0648&.0702&&.0729&.0623&.0697&&.0720&.0716&.0743\\
&&M=200&.0592&.0566&.0623&&.0593&.0611&.0645&&.0624&.0569&.0614\\
\hline
\end{tabular}
\end{center}
 \caption*{We  generated 10,000 independent sets of combined random samples. Each combined sample consists of a sample size of $N$ treated as diseased and a sample size of $N=r_a\times M$ treated as non-diseased. We discretized each sample into k categories, where k takes values k=3, k=7, and k=16.  The coverage probabilities reported in the above table are the fractions of the simulations falling outside the confidence interval $I^{c j}_{1-\alpha} $ described in Section~\ref{Variance Estimation}. Here $\alpha=0.05$.}
  \label{cover}
\end{table*}
\vspace{.4in}

\section{Application}
\label{application}

In Section \ref{pava} we showed how to obtain the constrained ROC curve
estimate using data from \citet{Chan2001}.
Figure~\ref{Chan unconstrained} shows the application of the 
PAVA to the ROC curve of radiologist~5 described in Section \ref{dataset}.
Ten ordinal categories were reduced to 6 categories to construct 
a convex ROC curve.  In this section we use the same data to compute
the constrained and unconstrained estimates of AUC as well as 
the estimates of variance of those AUC values.  All these 
estimates are given in Table~\ref{statistics}.  Also given are 
the results for radiologist~4 reading images at a different resolution.

\begin{table*}
\caption{Distribution of Score Ratings}
  \begin{center}
\begin{tabular}{|c|c|c|c|c|c|c|c|c|c|c|}
\hline Ordered Rating Category & 1 & 2 & 3 & 4 & 5 & 6 & 7 & 8 & 9 & 10 \\\hline
Diseased Counts ($n_i$) & 10 & 8 & 4 & 0 & 4 & 3 & 4 & 2 & 4 & 1 \\
Nondiseased Counts ($m_i$)& 33 &17 &6 &6 &5 &1 &0 &3 &1 & 0 \\ 
Likelihood Ratio ($\bar{w}_i$)& 0.30& 0.47& 0.67 &0.00& 0.80& 3.00 &$\infty$& 0.67& 4.00 &$\infty$ \\
 \hline
Constrained Diseased Counts ($\tilde{n_i}$) & 10 & \multicolumn{3}{c|}{12} & 4 &\multicolumn{3}{c|}{9} & 4 & 1 \\
Constrained Nondiseased Counts ($\tilde{m_i}$)& 33 & \multicolumn{3}{c|}{29} &5 & \multicolumn{3}{c|}{4} &1 & 0 \\ 
Likelihood Ratio ($\tilde{w_i}$)& 0.30& \multicolumn{3}{c|}{0.41}& 0.80& \multicolumn{3}{c|}{2.25} & 4.00 &$\infty$ \\ \hline
\end{tabular}
\end{center}
\caption*{This table gives the distribution of scores given by radiologist~5 
for diseased and nondiseased patients.  The lower rows show how adjacent
ordered rating categories are combined to obtain a monotonically increasing 
likelihood ratio.  This combined data is then used to construct an ROC curve 
and estimate AUC and its variance in the same manner  as the original data.
}
\label{reader5counttable}
\end{table*}

\begin{table}[h!]
  \begin{center}
\begin{tabular}{lllccr}

 &\multicolumn{2}{c}{Unconstrained}&&\multicolumn{2}{c}{Constrained}\\
&AUC&Variance&&AUC&Variance\\
\hline
Radiologist 4&.7386&.002605&&.7618& .002471\\
Radiologist 5&.6622&.002886&&.68472&.002782\\

\hline
\end{tabular}
\end{center}
  \caption{This table gives the constrained and unconstrained estimates 
of AUC and the variance AUC given given the ratings of radiologist 5 evaluating
lesions at a resolution of 140 microns. The same estimates are given for radiologist number 4 evaluating the lesions at a resolution
of 35 microns.
The estimates of AUC were calculated using Expressions~\ref{aucsum1} 
and~\ref{aucsum2}, and the variance was estimated using Expression~\ref{var}.
}
  \label{statistics}
\end{table}

As expected the constrained AUC estimates are greater than the 
unconstrained AUC estimates.  Correspondingly the variance estimates 
of the constrained AUC estimates are smaller than the unconstrained 
variance estimates.

\section{Summary}

In some studies of observer signal-detection performance 
we may assume that decisions made by human observers are rational,
implying the convexity of a population ROC curve.  In that scenario
we may want to analyze the data under that assumption.

In this paper we examined one such method, 
the constrained nonparametric maximum likelihood
ROC curve estimate given by 
\citet{lloyd2002}.  We proved directly that this 
this estimate is a simple convex hull around the usual empirical
ROC curve. 
We examined the properties of the area under this estimate
and found that it has lower variance than the usual empirical AUC estimate,
and it may also have lower bias with respect to a continous
nondiscretized ROC curve.
Our simulations indicate that 
standard estimators of  variance of the Mann-Whitney-Wilcoxon
statistic (AUC) variance may be used to obtain
nearly unbiased and very robust variance estimates of constrained 
nonparametric maximum likelihood AUC estimate.  
Bootstrapping of the convexity algorithm (PAVA) is not required.

The constrained ROC curve estimate is simple to construct in a 
manner similar to 
that suggested in \citet{Pesce2010}.
First a regular empirical ROC curve is constructed.  Then the lowest 
convex envelope that will enclose this ROC curve is drawn. 
This envelope is the new constrained nonparametric maximum likelihood 
ROC curve estimate.  Holding the total number
of observations constant, we recalculate the number of
observations in each category implied by the constrained estimate.
This is equivalent to combining adjacent categories that do not have 
increasing likelihood ratios, as in Table~\ref{reader5counttable}.
Next we use these new combined observation numbers 
to calculate the empirical AUC estimate in the usual manner.  This
is the constrained NPMLE of AUC.  We also can use those new combined
observations in standard nonparametric estimators of AUC variance 
to estimate the variance of the constrained NPMLE of AUC.
This estimator is a useful option for nonparametric analysis of data
from studies of observer signal-detection performance where the observers
are assumed to generate convex ROC curves rationally.

\section{Acknowledgements}
The authors are grateful to Dr. Chan and the other authors of 
Chan et al. 2001 for the use of their data in this paper.

\bibliographystyle{model5-names}
\bibliography{refs}

\end{document}